\documentclass[%
 reprint,
 amsmath,amssymb,
 aps,
]{revtex4-2}

\usepackage{graphicx}
\usepackage{dcolumn}
\usepackage{bm}

\usepackage{amsmath,amssymb}
\usepackage{array}
\usepackage{multirow}
\usepackage{graphicx}
\usepackage{hhline}
\usepackage[cdot,mediumqspace,amssymb]{SIunits}
\usepackage{xfrac}
\usepackage[normalem]{ulem}

\newcommand{\vect}[1]{\boldsymbol{#1}}

\begin{document}
\preprint{APS/123-QED}

\title{Vacuum polarization in molecules I: Uehling interaction}

\author{D. J. Flynn}
\affiliation{School of Physics, The University of Melbourne, Victoria 3010, Australia}

\author{I. P. Grant}
\affiliation{Mathematical Institute, University of Oxford, Oxford, United Kingdom}
\affiliation{Department of Applied Mathematics and Theoretical Physics, University of Cambridge, Cambridge, United Kingdom}

\author{H. M. Quiney}
\affiliation{School of Physics, The University of Melbourne, Victoria 3010, Australia}

\date{\today}

\begin{abstract}
    Radiative corrections to electronic structure are characterized by perturbative expansions in $\alpha$ and $Z\alpha$, where $\alpha$ is the fine-structure constant and $Z$ is the nuclear charge. A formulation of the leading-order $\alpha(Z\alpha)$ Uehling contribution to the renormalized vacuum polarization is reported in a form that is convenient for implementation in computational studies of the relativistic electronic structures of molecules. Benchmark calculations based on these methods are reported that include the leading-order vacuum polarization effects within relativistic mean-field methods for the E119$^+$ ion and the closed-shell diatomic species E119F.
\end{abstract}

\maketitle

\section{Introduction}\label{sec:intro}
Quantum electrodynamics (QED) unifies quantum mechanics and special relativity. It is a quantum field theory that describes the interactions between light and electrically charged particles, and the interactions between charged particles. This general description covers the physical phenomena encountered in atomic, molecular and solid-state physics and quantum chemistry. The theoretical and computational practices of these disciplines generally involve, however, approximations that can obscure their connections to the underlying foundations of QED. Many theoretical and computational approaches have been proposed to study the physics and chemistry of simple systems, such as atoms and small molecules, in order to more closely reflect these foundational principles. This work has been motivated by a desire to clearly establish the underlying theoretical foundations of complex systems, to explore the properties of the heavy and superheavy elements, to investigate phenomena at the interface of QED with electroweak theory, and to calibrate experiments that aim to detect physics beyond the Standard Model.\\

The electric field in the neighbourhood of a heavy nucleus induces a range of physical phenomena that require a more detailed description that is afforded by non-relativistic quantum theory. The desire to examine more closely the details of such effects serves as a motivation to the development of a complete computational implementation of QED. This includes a fully relativistic many-body formulation that incorporates radiative effects such as the electron self-energy and the polarization of the electron-positron vacuum.\\

Computational implementations of relativistic atomic physics have been under constant development for almost sixty years. This has been greatly facilitated by the adoption of Racah algebra techniques in the separation of radial and spin-angular variables in the construction of atomic state functions. The development of computer codes for atomic structures, such as \texttt{GRASP}~\cite{dyall1989grasp}, \texttt{QEDMOD}~\cite{shabaev2015qedmod} and \texttt{AMBiT}~\cite{kahl2019ambit}, has been driven by the demands of high-precision spectroscopy, and research applications that range from astrophysics to materials science. The extension of these methods to the calculation of molecular structures is complicated by a reduction in physical symmetry, which precludes the separation of single-particle solutions of the Dirac equation into simpler radial and spin-angular parts. A number of related approaches to this challenge have been developed into computational algorithms embodied in computer codes such as {\tt DIRAC}~\cite{saue2020dirac} and {\tt BERTHA}~\cite{FlynnGrantQuiney2023}. These codes adopt a finite basis set of square-integrable basis functions to represent the electronic amplitudes, and to facilitate the efficient evaluation of the matrix elements that arise in relativistic electronic structure theory. The details of how this is achieved in each code differ, but the general principles and the results are the same.\\

The conventional starting point for the relativistic electronic structure theory of atoms and molecules is the Dirac-Coulomb or Dirac-Coulomb-Breit hamiltonian, interpreted within a second-quantized Fock space formulation. The interactions between charged particles is represented by an effective interaction potential that contains contributions from both the charge and current densities. A mean-field approach defines the single-particle quanta of the field, cleanly divided into states of positive and negative energy. On this basis, one may construct a many-electron reference state, and relativistic many-body theory following the second-quantization conventions of QED. Many-body contributions involving negative-energy states are conventionally neglected, which is an approach usually described as the ``no-virtual-pair approximation" (NVPA).\\

The most significant corrections to this approximation are the electron self-energy and vacuum polarization effects, which in hydrogenic systems give rise to the Lamb shift. At leading order in perturbation theory, the calculation of the self-energy term is a formidable task requiring mass renormalization~\cite{grant2022grasp}. The structure of the effect is intrinsically non-local, but is often approximated by an effective local interaction to circumvent the computational challenges of an {\em ab initio} approach. The leading-order part of the vacuum polarization requires charge renormalization, but may be represented exactly by an local effective electrostatic potential, first derived by Uehling in 1935. The presence of the electrostatic field due to a nucleus causes the polarization of the electron-positron vacuum and the creation of virtual pairs. This mechanism causes a short-range modification of the classical electrostatic Coulomb potential due to the nuclear charge density, a shift in electronic amplitudes on length scales less than the Compton wavelength of the electron, $\lambdabar_{e} = 1/m_{e}c \approx 386 \; \text{fm}$, and a consequent shift in electronic energy levels.\\

The Uehling potential associated with a given spherically-symmetric nuclear charge density is readily evaluated as a numerical function of the radial distance from the centre of the nucleus. Despite the complexity of its derivation, it is simply an electrostatic potential that can be incorporated in electronic structure calculations, on the same footing as, for example, the nuclear Coulomb potential. For atoms, the Uehling potential is readily incorporated in electronic structure calculations if sufficient care is taken in choosing a radial grid for the numerical determination of the required matrix elements. For molecules, the cellular numerical integration scheme widely used in density functional theory may be adapted to evaluate the matrix elements of the Uehling potential~\cite{sunaga20224}. Gaussian fitting schemes that represent approximations to the Uehling potential have also been employed ~\cite{shepler2005ab,shepler2007hg}, which facilitate the evaluation of the required matrix elements in closed form.\\

This article reviews the theoretical and computational foundations of electronic structure theory, nuclear structure models and vacuum polarization in Section~\ref{formulation}. A Gaussian fitting scheme for the efficient representation of vacuum polarization phenomena is outlined in Section~\ref{gaussianfitting}. Numerical results are presented in Section~\ref{resultsdiscussion}, including comparisons with the atomic structure code {\tt GRASP} of the E119$^+$ ion and a study of the superheavy diatomic molecule E119F. An assessment of the significance of this work is included in the concluding remarks in Section~\ref{conclusion}.

\section{Formulation}\label{formulation}

\subsection{Relativistic electronic structure theory}
In our computational implementation of relativistic electronic structure theory, {\tt BERTHA}, every four-spinor solution of the one-electron Dirac equation, $\psi_k(\vect{r})$, is written as an expansion in a $G$-spinor basis set
\begin{equation}
	\psi_k(\vect{r}) = 
	\left[
	\begin{array}{l}
		\displaystyle		\sum_{\mu=1}^{N_B} c_{k,\mu}^L \mathcal{M}[L,\kappa_{\mu},m_{\mu},\vect{A}_{\mu},\lambda_{\mu}; \vect{r}] \\
		\displaystyle		i\sum_{\mu=1}^{N_B} c_{k,\mu}^S \mathcal{M}[S,\kappa_{\mu},m_{\mu},\vect{A}_{\mu},\lambda_{\mu};\vect{r}]
	\end{array}
	\right]
\end{equation}
where $N_B$ denotes the dimension of the basis set representation, $\mathcal{M}[T, \kappa_{\mu},m_{\mu},\vect{A}_{\mu},\lambda_{\mu};\vect{r}]$ is a two-component basis function characterized by the relativistic quantum numbers $\kappa_{\mu}$ and $m_{\mu}$, a Gaussian exponent, $\lambda_{\mu}$, and a local coordinate origin, $\vect{A}_{\mu}$. The component label $T=L$ corresponds to the so-called ``large" components, and $T=S$ the ``small'' components, which are useful designations for positive-energy bound-states. The same value of $N_B$ is used for $T=L$ and $T=S$ and the functions $\mathcal{M}[T, \kappa_{\mu},m_{\mu},\vect{A}_{\mu},\lambda_{\mu};\vect{r}]$ are generated by a mapping often described as ``kinetic balance". The detailed form of the $G$-spinor basis functions and the methods employed for evaluating relativistic matrix elements and implementing mean-field methods may be found in \cite{quiney1997,grant2006}.

\subsection{Nuclear structure}
A detailed survey of nuclear structure models has been provided by Andrae~\cite{andrae2000}. The nuclear models that are most widely used in relativistic electronic structure theory are the Gaussian charge model, the homogeneous ball model, and the two-parameter Fermi model. They are parametrized to match a specified root-mean-square (RMS) radius, $R$. For a nucleus of atomic mass $A$ this radius may be specified by the semi-empirical formula 
\begin{equation}
	R = (0.836A^{1/3} + 0.57) \; \text{fm},
\end{equation}
so that $R \approx 6.2762 \;$fm for an E119 nucleus of atomic mass $A = 318$. This is the superheavy nucleus that we employ in numerical calculations in Section~\ref{resultsdiscussion}.\\

Spherically symmetric nuclear model charge distributions, $\varrho(r)$, are related to the electrostatic potentials that they generate, $V(r)$, through the differential form of the Poisson equation, 
\begin{equation}\label{eq:poisson-forward}
	4 \pi r^{2} \varrho(r) = - r \tfrac{\partial^{2}}{\partial r^{2}} \bigl( rV(r) \bigr)
\end{equation}
or the equivalent integral form
\begin{equation}\label{eq:poisson}
	V(r) = -4 \pi \Bigl[ \tfrac{1}{r} \int_{0}^{r} s^{2} \varrho(s) \; \mathrm{d}s + \int_{r}^{\infty} s \varrho(s) \; \mathrm{d}s \Bigr].
\end{equation}
For a given nuclear model, the electric form factor, $G(q)$, is obtained from the Fourier transform of its charge density, $\varrho(r)$. For spherical nuclei, this relation is given by 
\begin{equation}
	G(q) = \int_{0}^{\infty} 4 \pi r^{2} \varrho(r) j_{0}(rq) \; \mathrm{d}r
\end{equation}
where $j_{n}(x)$ is the spherical Bessel function~\cite{abramowitz1970}. For small $q$ and $\varrho(r)$ normalized to unity, it is readily shown that 
\begin{equation}\notag
	G(q) = 1 - \tfrac{1}{6} q^{2} \langle r^{2} \rangle + \tfrac{1}{120} q^{4} \langle r^{4} \rangle - \cdots \label{eq:formfactor-powerseries}
\end{equation}
where the radial moments of the nuclear distribution, such as $\langle r^{2} \rangle=R^2$, are defined by
\begin{equation}
	\langle r^{k} \rangle
	= \int_{0}^{\infty} 4 \pi r^{k+2} \varrho(r) \; \mathrm{d}r.
\end{equation}
We specify a nuclear density located at position $\vect{A}_n$ by first noting that a multiplicative factor of $Z$ may later accommodate the total nuclear charge, and use the model distributions
\begin{eqnarray}\label{eq:gaussian-nucleus}
	\varrho_{G,n}(\vect{r}-\vect{A}_n) &=& \left(\tfrac{3}{2\pi R^2}\right)^{3/2}\exp\left(-\tfrac{3r_n^2}{2R^2}\right) \\	
	\varrho_{F,n}(\vect{r}-\vect{A}_n) \label{rhonucg} &=& \frac{\varrho_0}{1+\exp\left(\frac{r_n-a}{d}\right) } \\
	\varrho_{H,n}(\vect{r}-\vect{A}_n) &=& \frac{3}{4\pi {R_0}^3} \ \ \mbox{for $0\leq r_n\leq R_0$} \nonumber \\
	 & = & 0 \ \ \ \ \ \ \ \ \ 
	 \mbox{otherwise}
\end{eqnarray}
where $R$ is the mean-square nuclear radius, $R_0=\sqrt{5/3}R$ is the hard-sphere radius of the homogeneous model whose density is $\varrho_{H,n}(\vect{r}-\vect{A}_n)$, and $r_n=|\vect{r}-\vect{A}_n|$. The Gaussian distribution is denoted $\varrho_{G,n}(\vect{r}-\vect{A}_n)$, and incorporates the point nuclear distribution in the limit $R\to 0$. The Fermi nucleus, $\varrho_{F,n}(\vect{r}-\vect{A}_n)$, includes a parameter $a$ that is related to the effective nuclear skin thickness, $t$, through $a=t/4 \ln 3$; we use $t=2.3$~fm. The value of $\varrho_0$ is chosen so that $\varrho_{F,n}(\vect{r}-\vect{A}_n)$ represents a total charge of unity, and the other parameters are defined through the relations
\begin{eqnarray}
	d&=& \sqrt{\tfrac{5}{3}R^2-\tfrac{7}{3}\pi^2a^2}, \\
	\varrho_0&=& \tfrac{3}{4\pi d^3}\left[1+ \pi^2u^2 -6 u^3 S_3\left(-\tfrac{1}{u}\right)\right]^{-1}, \\
	S_k(x) &=& \sum_{n=1}^{\infty}(-1)^n\frac{\exp(nx)}{n^k},	
\end{eqnarray} 
and $u=a/d$.

\subsection{Vacuum polarization}
The approach we have adopted in the evaluation of vacuum polarization effects is summarized in Figure~\ref{fig:vp-connections}. It is based on the interrelations between the nuclear and polarization charge densities, $\varrho(r)$ and $\widetilde{\varrho}(r)$, respectively. They are each associated with corresponding potentials, $V(r)$ and $\widetilde{V}(r)$ through the Poisson equation, and to form factors $G(q)$ and $\widetilde{G}(q)$ through Fourier transform relations. This section investigates how the availability of a simple algebraic expression for the spectral function, $\Pi(\lambdabar q)$, of the Uehling potential for a structureless point nuclear distribution facilitates the development of complementary approaches to the calculation of vacuum polarization effects in relativistic electronic structure theories.

\subsubsection{Uehling potential}
For a structureless point nucleus of charge $Z$, the Uehling potential $\widetilde{V}^{(0)}(r)$ is given by
\begin{eqnarray}\notag
	\widetilde{V}^{(0)}(r)
	&=& -\tfrac{Z \alpha}{\pi} \tfrac{1}{r} \int_{1}^{\infty} \tfrac{2}{3t} \sqrt{1 - \tfrac{1}{t^{2}}} \bigl( 1 + \tfrac{1}{2t^{2}} \bigr) e^{-2tr/\lambdabar} \; \mathrm{d}t\\ \label{eq:uehling-potential-structureless}
	&=& -\tfrac{Z \alpha}{\pi r} \chi \bigl(\tfrac{2r}{\lambdabar};1\bigr),
\end{eqnarray}
where we define the auxiliary function, $\chi(x; k)$, to be
\begin{equation}\label{eq:chifunc}
	\chi (x; k) = \tfrac{2}{3} \int_{1}^{\infty} t^{-k} \sqrt{1 - \tfrac{1}{t^{2}}} \left( 1 + \tfrac{1}{2t^{2}} \right) e^{-xt} \; \mathrm{d}t.
\end{equation}
Chebyshev polynomial expansions of $\chi (x; k)$~\cite{fullerton1976} offer an efficient and accurate strategy for their evaluation. We have adopted methods based on an expansion in exponential integrals $E_{k}(x)$~\cite{glauber1960vacuum,huang1976calculation}. A cubic spline interpolation scheme on a grid of values permits the efficient and reliable calculation of the integrals on demand.\\

\begin{figure}[!b]
	\begin{center}
		\includegraphics[width=8.0cm]{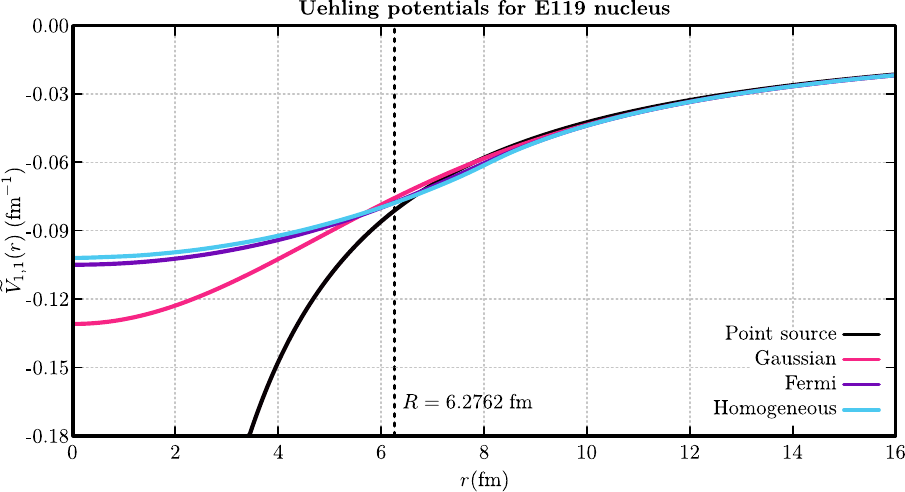}
		\caption{The Uehling potential for an E119 nucleus with radius $R = 6.2762 \;$fm, for Gaussian, Fermi and homogeneous nuclear charge models compared to the structureless case.}
		\label{fig:vp-E119-potential}
	\end{center}
\end{figure}

The Uehling potential, $\widetilde{V}(r)$, for a given finite nuclear charge density, $\varrho(r)$, can be obtained from the convolution theorem~\cite{fullerton1976}, yielding
\begin{eqnarray}\label{eq:uehling-potential}
	\widetilde{V}(r)
	&=& \varrho(r) \circledast \widetilde{V}^{(0)}(r) \\ \notag
	&=& -\tfrac{Z \alpha \lambdabar}{r} \int_{0}^{\infty} s \varrho(s) \Bigl[ \chi \bigl(\tfrac{2|r-s|}{\lambdabar} ; 2 \bigr) - \chi \bigl( \tfrac{2(r+s)}{\lambdabar} ; 2\bigr) \Bigr] \; \mathrm{d}s.
\end{eqnarray}
For the case of an E119 nucleus, we have selected a number of common nuclear charge models and presented their associated Uehling potentials in Fig.~\ref{fig:vp-E119-potential}. Of interest is their mutual convergence at distances $r \gtrsim 2R$.

\subsubsection{Polarization charge density}
The Uehling potential generated by any nuclear charge density may be thought of as being associated with a renormalized vacuum charge density, $\widetilde{\varrho}(r)$, induced in the vacuum by the generation of virtual electron-positron pairs. Poisson's equation may be expressed here as
\begin{eqnarray}\label{eq:polariseddensity}
    4 \pi r^{2} \widetilde{\varrho}(r) 
    &=& -r \tfrac{\partial^{2}}{\partial r^{2}} \bigl( r \widetilde{V}(r) \bigr) \\ \notag
    &=& \lambdabar Z \alpha r \int_{1}^{\infty} \tfrac{2}{3t^{2}} \sqrt{1 - \tfrac{1}{t^{2}}} \left( 1 + \tfrac{1}{2t^{2}} \right) Y(r,t) \; \mathrm{d}t
\end{eqnarray}
where
\begin{eqnarray*}
\lefteqn{Y(r,t) = -\tfrac{4t}{\lambdabar} r \varrho(r)}\\
&& + \tfrac{4t^{2}}{\lambdabar^{2}} \int_{0}^{\infty} s \varrho(s) \left( \exp \left[ -\tfrac{2|r-s|t}{\lambdabar} \right] - \exp \left[ -\tfrac{2(r+s)t}{\lambdabar} \right] \right) \; \mathrm{d}s.
\end{eqnarray*}
One may immediately state that the three-dimensional volume integral over $\widetilde{\varrho}(\vect{r})$ must vanish identically, because the induced charges are created as virtual pairs. Despite the formidable algebraic complexity of Eq.~(\ref{eq:polariseddensity}), we have confirmed by numerical quadrature that this fundamental property is satisfied.\\

\begin{figure}[!h]
	\begin{center}
		\includegraphics[width=8.0cm]{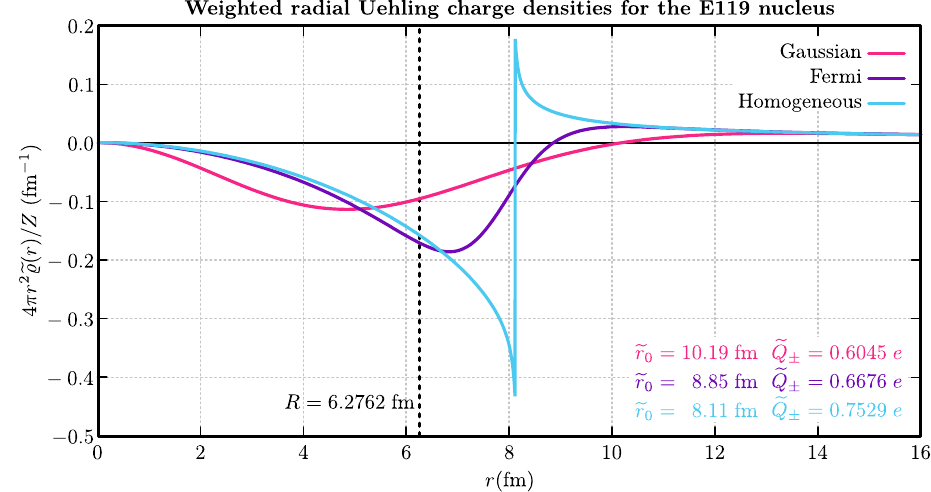}
		\caption{The weighted radial charge density for Gaussian, Fermi and homogeneous models of an E119 nucleus. The quantity $\widetilde{r}_0$ denotes the value of the radial coordinate for which the radially weighted charge density changes sign. The corresponding values $\widetilde{Q}_{\pm}$ indicate the magnitude of the charge enclosed inside and outside a sphere of radius $\widetilde{r}_0$.}
		\label{fig:vp-E119-density}
	\end{center}
\end{figure}

Figure~\ref{fig:vp-E119-density} compares the vacuum charge densities for Gaussian, Fermi and homogeneous models of the E119 nucleus. The quantity $\widetilde{r}_0$ in Figure~\ref{fig:vp-E119-density} denotes the value of the radial coordinate for which the radially weighted charge density changes sign. The corresponding values $\widetilde{Q}_{\pm}$ indicate the magnitude of the charge enclosed inside and outside a sphere of radius $\widetilde{r}_0$, confirming that the net vacuum charge vanishes identically.
\vspace{-0.9cm}

\subsubsection{Polarization spectral function}\label{subsubsec:spectral}
An alternate perspective on vacuum polarisation can be sought using the momentum space representation. A Fourier transform relation,
\begin{equation}\label{eq:ft-spectral-to-potential}
	\widetilde{V}^{(0)}(r)
	= \mathcal{F}^{-1} \bigl[ - \tfrac{4\pi}{q^{2}} \Pi(\lambdabar q) \bigr],
\end{equation}
may be defined between $\widetilde{V}^{(0)}(r)$ and a spectral function, $\Pi(\lambdabar q)$, where $\lambdabar$ is the Compton wavelength. It is convenient to define the unitless parameter $u = \lambdabar q$ to describe the point-charge spectral function for any virtual fermionic field 
\begin{equation}\label{eq:spectral-function}
	\Pi(u) = \tfrac{Z \alpha}{\pi} \bigl( - \tfrac{5}{6} + \tfrac{2}{u^{2}} + \tfrac{u^{2}-2}{u^{3}} \sqrt{u^{2}+4} \; \text{arcsinh}  (\tfrac{u}{2}) \bigr).
\end{equation}

\begin{figure}[!h]
	\begin{center}
		\includegraphics[width=8.0cm]{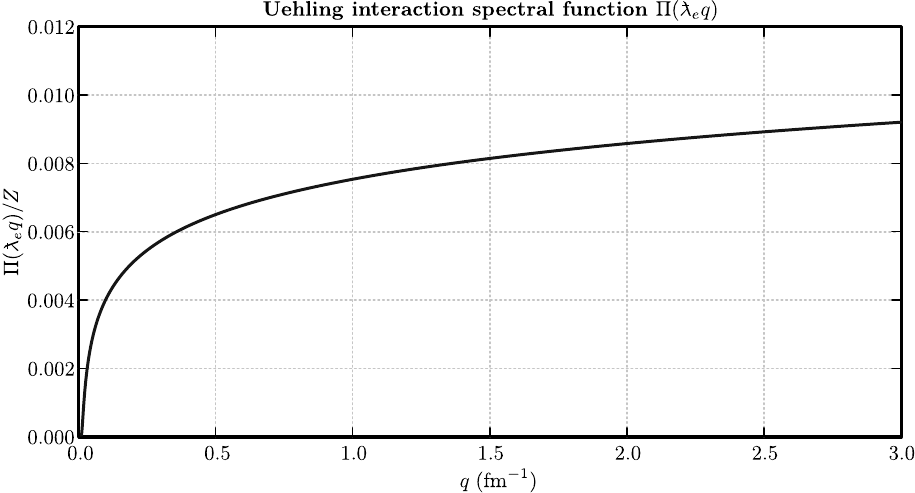}
		\caption{The spectral function which accounts for the Uehling interaction, in its coupling to the electron-positron field for which the Compton wavelength is $\lambdabar_e$.}
		\label{fig:vp-uehling-spectral}
	\end{center}
\end{figure}

In the case of the electron-positron field where $\lambdabar = \lambdabar_{e} \approx 386 \; \text{fm}$, the spectral function has the form indicated in Fig.~\ref{fig:vp-uehling-spectral}.

In the low-$q$ region, $\Pi(u)$ may be represented by a power series expansion of the form
\begin{equation}\label{eq:spectral-taylor}
	\Pi(u)
	\to \tfrac{Z \alpha}{\pi} \bigl( -\tfrac{u^{2}}{10} + \tfrac{3u^{4}}{280} - \tfrac{u^{6}}{630} + \tfrac{u^{8}}{3696} - \cdots \bigr).
\end{equation}
In the ultra-relativistic limit, $q\to\infty$, one may employ the asymptotic expansion
\begin{equation}\label{eq:spectral-asymptotic}
	\Pi (u) \to \tfrac{Z \alpha}{\pi} \bigl(- \tfrac{5}{6} + \log (u) + \tfrac{3}{u^{2}} - \tfrac{3}{2u^{4}} + \tfrac{6}{u^{4}} \log (u) \bigr).
\end{equation}

\begin{figure}[!h]
	\begin{center}
		\includegraphics[width=8.0cm]{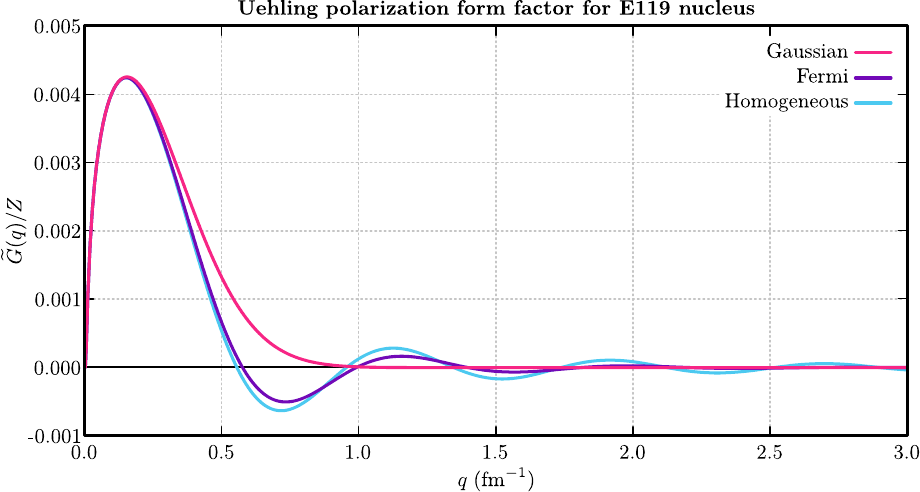}
		\caption{Polarization form factors for the Uehling interaction.}
		\label{fig:vp-E119-induced-formfactors}
	\end{center}
\end{figure}

For a normalized point nucleus, $\varrho(\vect{r}) = \delta^{(3)}(\vect{r})$, and its Fourier transform is $G(q) = 1$. The spectral function in Eq.~(\ref{eq:spectral-function}) is then precisely the form factor the describes the polarized vacuum arising from this nucleus. More generally, we may write
\begin{equation}\label{eq:polarisation-formfactor}
	\widetilde{G}(q) = \Pi(\lambdabar q) G(q).
\end{equation}
Provided that $G(q)$ is sufficiently smooth, we find that the polarization form factor is strongly attenuated for large $q$, as can be seen in Fig.~\ref{fig:vp-E119-induced-formfactors}. This is most easily verified for a Gaussian nuclear profile, for which $G_{g}(q) = e^{-q^{2}R^{2}/6}$. The spectral function for the electron-positron field is suppressed by a factor of $100$ when $q \approx 5/R$, which for a typical heavy nucleus corresponds to $q\approx 1~\text{fm}^{-1}$.\\

Details about the induced form factor near $q \approx 0$ embed some useful information about the underlying polarized vacuum. By combining the series expansion for the nuclear form factor in Eq.~(\ref{eq:formfactor-powerseries}) with the Taylor series expansion of the spectral function in Eq.~(\ref{eq:spectral-taylor}), and collecting terms in powers of $q$, we find that
\begin{eqnarray} \notag 
	\widetilde{G}(q)
	&=& G(q) \Pi(\lambdabar q) \\ \label{eq:induced-form-factor} 
	&=& \tfrac{Z \alpha}{\pi} \Bigl[ -\bigl( \tfrac{1}{10} \bigr) \lambdabar^{2} q^{2} + \bigl( \tfrac{3}{280} + \tfrac{1}{60} \tfrac{R^{2}}{\lambdabar^{2}} \bigr) \lambdabar^{4} q^{4} \\ \notag
	& & \qquad - \bigl( \tfrac{1}{630} + \tfrac{1}{560}\tfrac{R^{2}}{\lambdabar^{2}} + \tfrac{1}{1200}\tfrac{\langle r^{4} \rangle}{\lambdabar^{4}} \bigr) \lambdabar^{6} q^{6} + \cdots \Bigr].
\end{eqnarray}
The availability of a closed-form expression for $\Pi(u)$ and the ease with which $G(q)$ may be obtained for realistic nuclei conveys a clear advantage in the evaluation of quantities associated with vacuum polarization phenomena. This approach to the calculation of the effects of leading-order vacuum polarization is indicated by the green path in Fig.~\ref{fig:vp-connections}.\\

The function $\widetilde{G}(q)$ may also be expressed as an expansion in the radial moments of the vacuum charge density, $\langle \widetilde{r}^k \rangle$, 
\begin{eqnarray}\label{eq:inducedformfactor_powerseries}
	\widetilde{G}(q)
	&=& \int_{0}^{\infty} 4 \pi r^{2} \widetilde{\varrho}(r) j_{0}(rq) \; \mathrm{d}r \\ \notag
	&=&  Z \bigl(\langle \widetilde{r}^{0} \rangle - \tfrac{1}{6} q^{2} \langle \widetilde{r}^{2} \rangle + \tfrac{1}{120} q^{4} \langle \widetilde{r}^{4} \rangle - \tfrac{1}{5040} q^{6} \langle \widetilde{r}^{6} \rangle + \cdots \bigr).
\end{eqnarray}
By equating coefficients of Eq.~(\ref{eq:induced-form-factor}) and Eq.~(\ref{eq:inducedformfactor_powerseries}), we identify the even radial moments of the vacuum charge density as
\begin{subequations}\label{eq:polarisation-even-moments}
	\begin{eqnarray}\label{eq:r0}
		\langle \widetilde{r}^{0} \rangle
		&=& 0 \\ \label{eq:r2}
		\langle \widetilde{r}^{2} \rangle
		&=& \tfrac{2\alpha\lambdabar^{2}}{5 \pi} \\ \label{eq:r4}
		\langle \widetilde{r}^{4} \rangle
		&=& \tfrac{6 \alpha \lambdabar^{4}}{7 \pi} \bigl( 1 + \tfrac{14R^{2}}{9 \lambdabar^{2}} \bigr) \\ \label{eq:r6}
		\langle \widetilde{r}^{6} \rangle
		&=& \tfrac{16 \alpha \lambdabar^{6}}{3 \pi} \bigl( 1 + \tfrac{9R^{2}}{8 \lambdabar^{2}} + \tfrac{21\langle r^{4} \rangle}{40 \lambdabar^{4}} \bigr).
	\end{eqnarray}
\end{subequations}
The first of these relations, $\langle \widetilde{r}^0 \rangle=0$, is a consequence of the Ward identity~\cite{ward1950}, and the order-by-order conservation of charge; the net charge vanishes, because the charges are induced as virtual electron-positron pairs. The second relation, Eq.~(\ref{eq:r2}), indicates that the RMS radius of the induced density is determined solely by a combination of mathematical and physical constants, independent of the nuclear model. For the electron-positron field, $\widetilde{R}=\langle \widetilde{r}^2\rangle^{1/2}$ has the value $\widetilde{R} \approx 11.81 \;$fm, which is approximately twice the radius of a typical heavy nucleus. Details of the nuclear structure enter only in the definitions of $\langle\widetilde{r}^4 \rangle$ and the higher-order radial moments. These nuclear structure effects are evidently rather small for the electron-positron field because $R \ll \lambdabar$ for any real nucleus.
\begin{widetext}
	\begin{center}
		\begin{figure}[h]
			\includegraphics[width=14.0cm]{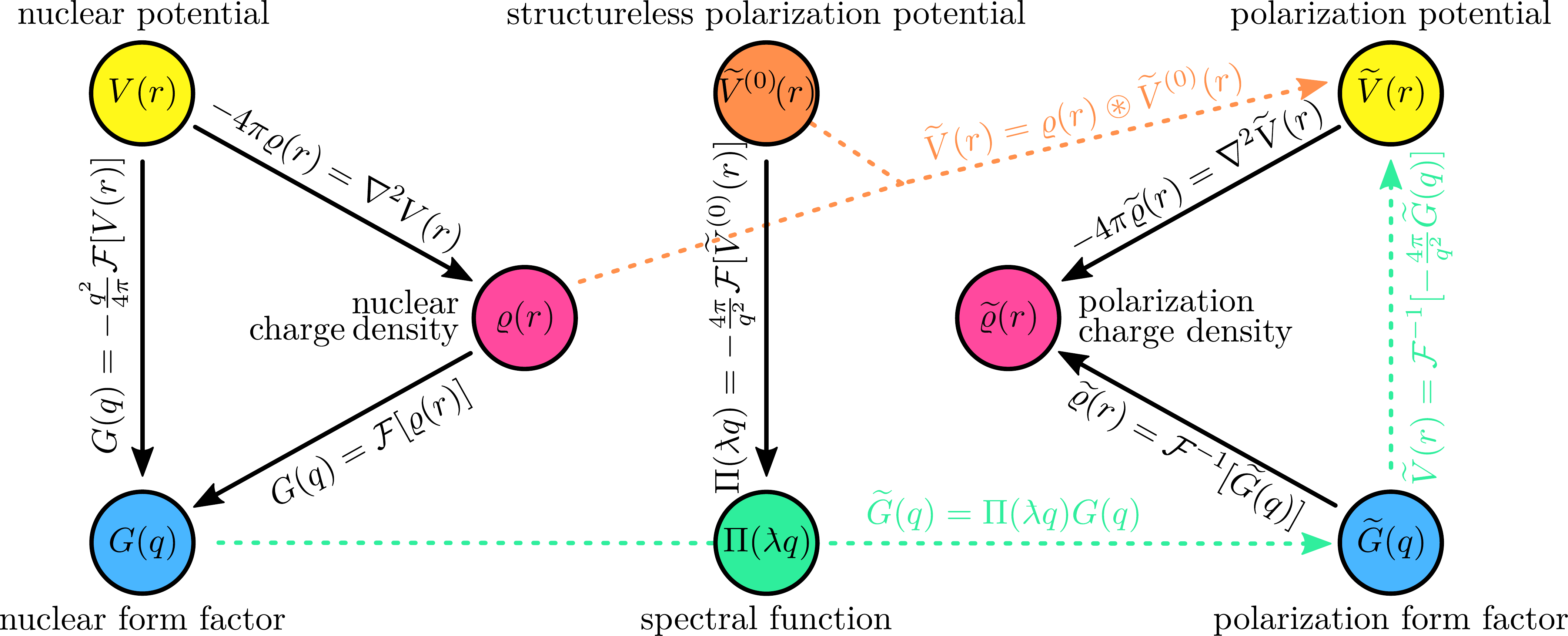}
			\caption{Connections between the key quantities in this computational study of vacuum polarization in atoms and molecules. Rather than calculating polarization potentials for realistic nuclear models using Eq.~(\ref{eq:uehling-potential}) (orange path), we employ the spectral function and Fourier transform relations, using Eq.~(\ref{eq:polarisation-formfactor}) and Eq.~(\ref{eq:ft-spectral-to-potential}) (green path).}
			\label{fig:vp-connections}
		\end{figure}
	\end{center}
\end{widetext}

\section{Gaussian fitting of the vacuum polarization}\label{gaussianfitting}
In relativistic electronic structure calculations, it has become common practice to employ a Gaussian charge density to include the effects of finite nuclear structure. This facilitates the inclusion of the associated nuclear electrostatic potential in relativistic matrix elements. Two important features of the Uehling potential are its short spatial range from the nucleus and that it is associated with a net polarization charge that vanishes identically. These properties suggest a Gaussian expansion of the Uehling potential of the form
\begin{equation}
	\widetilde{V}(r) = Z\sum_{n=1}^{N} A_{n} \exp [-\zeta_{n}r^{2}].
	\label{eq:fitvacuumpotential}
\end{equation}
The application of the Poisson equation generates an expansion of the vacuum charge density,
\begin{eqnarray}\label{eq:fitvacuumdensity}
	\widetilde{\varrho}(r) &=& -\tfrac{Z}{2 \pi} \sum_{n=1}^{N} A_{n}\zeta_{n} (3 - 2\zeta_{n}r^{2}) \exp [-\zeta_{n}r^{2}] \\
	&=& \sum_{n=1}^N A_n \xi_n(r),
\end{eqnarray}
in which the vanishing of net vacuum charge density is imposed term by term. For convenience, we refer to the functions $\xi_n(r)$ as ``zero charge Gaussian functions" because of this property, which follows from Gauss's law of electrostatics. The Fourier transform of $\widetilde{\varrho}(r)$ generates the corresponding Gaussian expansion for $\widetilde{G}(q)$ of the form 
\begin{equation}
	\widetilde{G}(q) = -\tfrac{Z\sqrt{\pi}}{4} q^{2} \sum_{n=1}^{N} A_{n} \zeta_{n}^{-3/2} \exp [-q^{2}/4 \zeta_{n}].
	\label{eq:fitvacuummomentum}
\end{equation}
The fitting of the vacuum polarization involves the specification of $N$ exponential parameters, $\zeta_n$, and the determination of $N$ associated expansion coefficients $A_n$. We have found that $N=26$ generates a sufficiently accurate expansion, subject to the imposition of the following constraints:
\renewcommand{\labelenumi}{\roman{enumi}}
\begin{enumerate}
	\item $\displaystyle \langle \widetilde{r}^2 \rangle
	= \tfrac{2 \alpha \lambdabar^{2}}{5 \pi}  =\tfrac{3 \sqrt{\pi}}{2} \sum_{n=1}^{N} A_{n} \zeta_{n}^{-3/2}$
	\item 
	$\displaystyle \langle \widetilde{r}^4 \rangle= \tfrac{6 \alpha \lambdabar^{4}}{7 \pi} \bigl( 1 + \tfrac{14R^{2}}{9 \lambdabar^{2}} \bigr)=\tfrac{15 \sqrt{\pi}}{2} \sum_{n=1}^{N} A_{n} \zeta_{n}^{-5/2}$
	\item $\displaystyle \widetilde{V}(r_i) = Z\sum_{n=1}^{N} A_{n} \exp [-\zeta_{n}r^{2}_i],\ i=1,2,\ldots,7 $
	\item $\displaystyle r^2_i\widetilde{V}(r_i) = Z\sum_{n=1}^{N} A_{n} r^2_i \exp [-\zeta_{n}r^{2}_i],\ i=8,9,\ldots,24. $
\end{enumerate}
The radial sample points for $1\leq i \leq 7$ are specified on a linear grid for $0\leq r_i \leq 2R$ and the points $8\leq i\leq 24$ are distributed on an exponential grid for $2R < r_i \leq 240R$.\\

We have adopted the well-tempered series to specify the exponential parameters, according to the prescription
\begin{equation}
	\zeta_{1} = \alpha, \quad
	\zeta_{n+1} = \zeta_{n} \beta \Bigl[ 1 + \gamma \bigl( \tfrac{n}{N+1} \bigr)^{2} \Bigr].
\end{equation}
where $\alpha$, $\beta$ and $\gamma$ are parameters. In order to limit the effects of linear dependence in the basis set, we have adopted the values $\beta=1.6$ and $\gamma=0.21$. The value of $\alpha$ determines the most spatially diffuse basis function, and we tune its value by performing a search on values in the range $2\times 10^{-5}\leq \alpha R^2 \leq 8\times 10^{-5}$. The chosen value of $\alpha$ is that which minimizes the absolute difference between the fitted potential and the true Uehling potential in a least-square sense.\\

The values of the fitting parameters $A_n$ are determined by solving the linear system $\vect{A}\vect{X}=\vect{C}$, where the elements of the square matrix $\vect{X}$ define the constraint equations and the elements of the vector $\vect{C}$ are the constraint values. We have included fitting parameters and Gaussian exponents for the E119 and F nuclei in the supplementary material.
\begin{figure}[!h]
	\begin{center}
		\includegraphics[width=8.0cm]{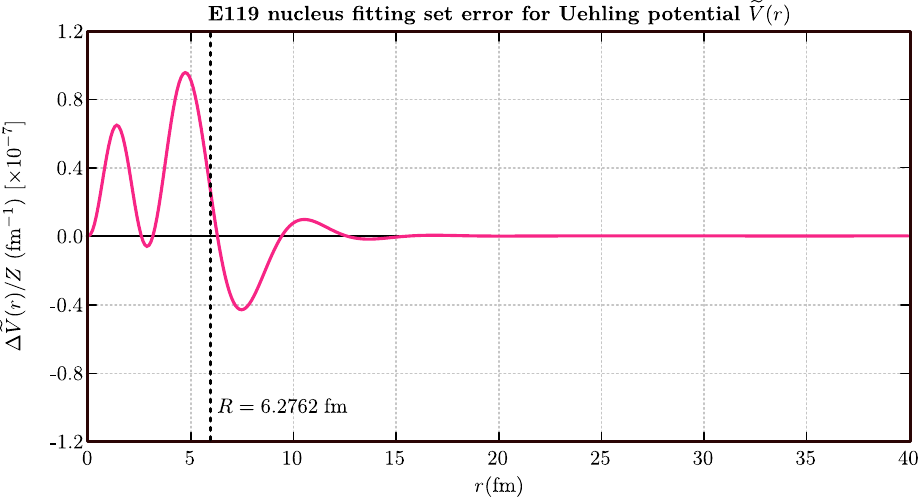}
		\caption{Fitting set error for the Uehling potential which arises from an E119 nucleus (Gaussian charge model), as it compares to a numerical evaluation.}
		\label{fig:uehling_fitting_difference}
	\end{center}
\end{figure}

The error in our fitting set for Uehling potential which arises from a Gaussian model of an E119 nucleus is displayed in Fig.~\ref{fig:uehling_fitting_difference}. The most marked pointwise discrepancies occur in the region $r \leq 2R$, resulting in at worst a 0.01\% pointwise fitting set error around $r \approx 5 \;\text{fm}$. For these small values of $r$, the fitting error is strongly suppressed by the radial factor of $r^2$ in the volume element that is included in the evaluation of matrix elements of the Uehling potential.

\section{Results and discussion}\label{resultsdiscussion}
In order to test our methods in systems for which the effects of vacuum polarization are large, we have selected test systems based on element 119, which we denote by E119. It is also known as ununennium or eka-francium and its synthesis is the subject of ongoing research in nuclear physics. The relativistic electronic structure calculations reported here were performed with {\tt BERTHA} and, where appropriate, compared with corresponding calculations generated by {\tt GRASP}. All matrix elements in {\tt BERTHA} are obtained by expanding the charge-current densities of the $G$-spinor basis sets in an auxiliary basis set of Gaussian type. Since the Uehling potential is expanded in Gaussian basis functions following Eq.~(\ref{eq:fitvacuumpotential}), all matrix elements may be evaluated using convenient algebraic expressions that are commonly encountered in quantum chemistry.

\subsection{E119$^+$ ion}
For our benchmark calculations using {\tt BERTHA}, we employ a very large $G$-spinor basis of size $75s 70p 65d 60f$. The Gaussian exponential parameters for all symmetry types are generated using a geometric series $\lambda_{n} = \alpha \beta^{n-1}$ with $\alpha = 0.00845$ and $\beta = 1.55$. The basis functions with largest values of the parameters $\lambda$ are sufficiently narrow that they significantly penetrate the nucleus, and are able to represent the short-range details of the spinors that are sensitive to the effects of vacuum polarization.\\

A Gaussian nuclear model, Eq.~(\ref{eq:gaussian-nucleus}), was employed for these calculations, using $R = 6.2762 \;$fm. The mean-field calculations were performed at three levels of complexity. The model we label as {\tt DHFR} includes only the Coulomb interaction in the mean-field potential, so that the effects of the Breit and Uehling interactions are evaluated by first-order perturbation theory. The {\tt DHFB} model also includes the Breit interaction in the definition of the mean-field potential, so that only the Uehling potential is evaluated by perturbation theory. Finally, the {\tt DHFQ} model includes the Coulomb, Breit and Uehling interactions variationally and self-consistently.\\

In order to assess the accuracy of the {\tt DHFR} calculations, we compare them with those obtained using {\tt GRASP} using the same nuclear model. Table~\ref{tab:E119+_energies_methods} records the Dirac-Hartree-Fock energies of the E119$^+$ ion calculated using the two programs, as well as the first-order expectation values of the Breit and Uehling interactions. In both cases, the expectation value of the Uehling interaction was calculated by numerical quadrature, using a tabulation of the radial electron density and the known form of $\widetilde{V}(r)$. The close agreement between the two calculations establishes the equivalence of the two codes, which are based on quite different numerical methods for solving the external-field Dirac equation. It also demonstrates the quality of the wavefunction generated by {\tt BERTHA}, compared to the standard provided by {\tt GRASP}.\\

\begin{table}[h!]
	\begin{center}
        \caption{\label{tab:E119+_energies_methods} Dirac-Hartree-Fock (DHF), Breit and Uehling energies (a.u.) and differences in energy, $\Delta E$, comparing {\tt GRASP} and {\tt BERTHA} for an E119$^{+}$ ion with a Gaussian nuclear charge. The Breit and Uehling interaction energies are evaluated using first-order perturbation theory.}
		\begin{tabular}{ l | r r | r } 
			\hline \hline
			& \texttt{GRASP}         & \texttt{BERTHA}              & $\Delta E$ \\ \hline
			DHF   & --56145.30016765                & --56145.30016777                & --0.00000012 \\
			Breit         & {102.63126999}  & {102.63126974}  & {--0.00000025} \\
			Uehling       & {--50.20593584} & {--50.20593583} & {  0.00000001} \\ \hline
			Total         & --56092.87483350                & --56092.87483384                & --0.00000036 \\
			\hline \hline
		\end{tabular}
	\end{center}
\end{table}

Having established that the wavefunction for E119$^+$ is of comparable quality for calculations using {\tt GRASP} and {\tt BERTHA}, we now test the accuracy of the fitted representation of $\widetilde{V}(r)$. Table~\ref{tab:E119+_energies_hamiltonians} records the one-electron, Coulomb, Breit and Uehling energies in the {\tt DHFR}, {\tt DHFB} and {\tt DHFQ} electronic models using a Gaussian nuclear model within {\tt BERTHA}. In particular, we find that the error introduced by adopting a Gaussian expansion of the Uehling potential is only $1.6\times 10^{-7}$ a.u. in E119$^+$, compared with the value $-50.20593584$~a.u., which is the benchmark Uehling energy obtained by {\tt GRASP}.

\begin{table}[h!]
	\begin{center}
        \caption{\label{tab:E119+_energies_hamiltonians} One-body, Coulomb, Breit and Uehling energy contributions (a.u.) to the electronic energy of the E119$^{+}$ ion, calculated using the {\tt DHFR}, {\tt DHFB} and {\tt DHFQ} electronic models and a Gaussian nuclear model. The Breit and Uehling energies in the {\tt DHFR} model and the Uehling energy in the {\tt DHFB} model are calculated by first-order perturbation theory. The other quantities are calculated as contributions to the Coulomb, Breit and Uehling energies in the corresponding mean-field models.}
		\begin{tabular}{ l | r r r } 
			\hline \hline
			& \texttt{DHFR}    & \texttt{DHFB}        & \texttt{DHFQ}   \\ \hline
			One-body   & --74814.85434779 & --74785.30449729 & --74788.70402623 \\
			Coulomb &   18669.55418002 &   18640.22401196 &   18643.67943181 \\
			Breit   & {102.63126974} & 102.19144558 & 102.44652427 \\
			Uehling & {--50.20593567} & {--49.95173100} & --50.57134230 \\ \hline
			Total   & --56092.87483370 & --56092.84077075 & --56093.14941245 \\
			\hline \hline
		\end{tabular}
	\end{center}
\end{table}

It has long been established that the total energies calculated in the {\tt DHFR} and {\tt DHFB} models are always very similar, if the Breit interaction is included as a perturbation to the {\tt DHFR} model. In this superheavy element species, the difference between these two approaches is only 0.22 a.u. The effect of the Breit interaction is to increase the energy, and its self-consistent inclusion in the {\tt DHFB} model has the secondary effect of reducing the magnitude of the first-order expectation value of the Uehling energy, compared with the {\tt DHFR} model. The self-consistent treatment of the Uehling interaction in the {\tt DHFQ} model increases the magnitudes of all of the contributions to the total energy, especially the one-electron and Coulomb terms. Relative to the {\tt DHFB} model, these two terms are increased in magnitude by more than 3~a.u. When all contributions are included, however, the self-consistent treatment of the Uehling interaction decreases the total energy by only 0.31 a.u., relative to the {\tt DHFB} model.\\

We have also evaluated the second and- third-order perturbative Uehling corrections to the {\tt DHFB} model; these are, respectively, $E^{(2)}=-0.30790882$~a.u. and $E^{(3)}=-0.00230352$~a.u, so these higher-order corrections to the Uehling model sum approximately to $-0.31$~a.u. This is in excellent agreement with the difference in total energies for the {\tt DHFB} model and the variational inclusion of the Uehling interaction within the {\tt DHFQ} model. These additional terms correspond to ladder diagrams, which are summed through all orders when the \texttt{DHFQ} potential is used.

\subsection{E119F molecule}

The representation of $\widetilde{V}(\vect{r}-\vect{A}_n)$ in an auxiliary Gaussian basis set, Eq.~(\ref{eq:fitvacuumpotential}), for each nucleus labelled $n$, enables the evaluation of the $G$-spinor matrix elements of the Uehling interaction to be reduced to three-centre Gaussian overlap integrals, which are readily calculated using the existing integral technology within {\tt BERTHA}.\\

For this system, we adopt an energy-optimized basis set that is representative of the type widely used in relativistic quantum chemistry. For the E119 atomic centre we have selected the $34s31p23d18f$ basis set of Miranda \text{et al.}~\cite{miranda2012}. We have also used Dunning's optimized aug-cc-pVTZ basis set parameters for atomic fluorine~\cite{dunning1989}, resulting in a $18s 10p 3d$ set. The parameters for the Gaussian model of the E119 nucleus are the same as for E119$^{+}$, while we have used a Gaussian charge model for the fluorine centre with $A=18.99$ and $R = 2.8004$ fm. The internuclear separation has been chosen to be $d=2.448$~\AA~\cite{miranda2012}.\\

The decomposition of the total energy for the fragment ions E119$^{+}$ and F$^-$ into one-body, Coulomb, Breit and Uehling contributions using the {\tt DHFQ} electronic model and the Gaussian nuclear structure model is presented in Table~\ref{tab:E119F_breakdown}. The corresponding quantities for the diatomic molecule E119F, together with the nuclear repulsion energy, are presented in the same table. A number of striking features are apparent from these data. The formation of a molecular bond in E119F in this mean-field model causes a substantial redistribution of the one-body and Coulomb energies, involving several hundred a.u., relative to the values for the isolated fragment ions. Including the nuclear repulsion energy, however, leads to a net energetic stabilization of the molecule in this approximation of only 0.211~a.u. Comparing the Breit and Uehling contributions in the molecular and ionic fragment calculations, we find that these agree to the conventional measure of chemical accuracy, 0.001~a.u.\\

The decomposition of the Breit and Uehling energies into one- and multi-centre contributions demonstrates that these two effects involve multi-centre contributions with magnitudes of approximately 0.0001~a.u. For a closed-shell system, the Breit interaction only involves exchange contributions, which suppresses multi-centre effects. For both the Breit and Uehling interactions, the contributions are dominated by inner-shell spinors, so that a one-centre approximation appears to be sufficient for most practical purposes. The self-consistent treatment of the Uehling interaction appears to be a more important effect on the total energy than the inclusion of multi-centre contributions from either the Breit or Uehling interactions. The use of a fitted representation of the Uehling potential renders its self-consistent inclusion in mean-field approximations essentially trivial. The use of Racah algebra methods for the calculation of the one-centre Breit contributions~\cite{FlynnGrantQuiney2023} offers a particularly efficient computational scheme for the implementation of the {\tt DHFQ} electronic structure model. Together, these technical developments represent a significant practical advance towards the implementation of a comprehensive treatment of relativistic quantum electrodynamics in molecules. The additional computation time required to handle these interactions is typically less than 10\% of the time required for the \texttt{DHFR} calculation.

\begin{table*}
    \caption{\label{tab:E119F_breakdown} Energy expectation values (a.u.) involving the \texttt{DHFQ} hamiltonian for the E119F molecule, using basis sets from Miranda \text{et al.}~\cite{miranda2012} and Dunning~\cite{dunning1989} for E119$^{+}$ and F$^{-}$ respectively. \textsc{Ionic fragments}: The ionic constituents of the E119F molecule, generated from atomic basis set calculations. \textsc{Molecular contributions}: Decomposition into one- and multi-centre contributions from a molecular calculation, using a bond length $d = 2.448 \angstrom$. The Breit and vacuum polarization interactions lead to multi-centre contributions that are small compared to one-centre terms.}
    \begin{ruledtabular}
    \begin{tabular}{ l | r r | r || r r | r}
        & \multicolumn{3}{c||}{\textsc{Ionic fragments}} & \multicolumn{3}{c}{\textsc{Molecular contributions}}\\ \hline
        &       E119$^{+}$ &        F$^{-}$ &     Fragment sum & One-centre       & Multi-centre   & Molecule         \\ \hline
        Nuclei                     &               -- &             -- &               -- &               -- &   231.51502982 &     231.51502982 \\
        Dirac                      & --74788.69535572 & --144.11974256 & --74932.81509828 & --74954.10257055 & --464.35583406 & --75418.45840461 \\
        Coulomb                    &   18643.67152577 &    44.57475181 &   18688.24627758 &   18709.04130422 &   233.12178904 &   18942.16309326 \\
        Breit                      &     102.44651846 &     0.01140572 &     102.45792418 &     102.45801723 &   --0.00010835 &     102.45812558 \\
        Uehling ($e^{+}e^{-}$)     &    --50.57137187 &   --0.00038832 &    --50.57176019 &    --50.57195654 &   --0.00009114 &    --50.57204768
        \\ \hline
        Total                      & --56093.14868336 &  --99.53397335 & --56192.68265671  & --56193.17520564 &     0.28100202 & --56192.89420363 \\
        \end{tabular}
    \end{ruledtabular}
\end{table*}

\section{Conclusion}\label{conclusion}
A comprehensive treatment of the relativistic electronic structure of molecules containing heavy elements requires several computational components. The Coulomb, Breit and vacuum polarization contributions may all be included as effective potentials that are absorbed into the definition of the self-consistent field. The self-energy involves a non-local interaction, but may also be included in the {\tt DHFQ} model within a local approximation. The expansion of Dirac spinors in a finite basis set generates a discrete set of states that, in principle, may be extended systematically towards completeness. Inclusion of all significant one-body effective potentials in the self-consistent field procedure generates a set of discrete single-particle quanta that simplifies the subsequent treatment of many-body interactions. Brillouin's theorem applies equally as well to the relativistic theory as it does to the non-relativistic theory; one-body ``bubble diagram" corrections vanish to all orders in perturbation theory for potentials that are included self-consistently. As a consequence the many-body refinement of the calculation includes only two-body terms, including the residual parts of the Coulomb, Breit and frequency-dependent transverse interactions. The evaluation of the four-component spinor amplitudes is complicated in molecular studies by the need to evaluate multi-centre integrals involving operators that are not encountered in conventional quantum chemistry programs. These include the Breit interaction, frequency-dependent interaction corrections, the Uehling potential, local representations of the self-energy operator or higher-order QED interactions. These present technical problems that can be solved in a number of ways; we have chosen to develop a technology for {\tt BERTHA} that is based on the explicit representation of charge-current densities, so that the computational methods closely follow the underlying foundations in QED. The computer code {\tt DIRAC} has adapted the scalar technologies of existing quantum chemistry computer codes, but the two approaches are formally equivalent to within a linear transformation of their basis sets. A third approach seeks to reduce the computational cost by reducing the representations to scalar or two-component forms. This can be achieved, but at the expense of greatly complicating the relationship between the computational representation and the underlying foundations in quantum field theory.\\

We have shown that the Uehling potential may be fitted to high accuracy using an auxiliary Gaussian basis set that imposes physical constraints derived from the underlying QED theory. The zero-charge Gaussians that we have utilized embody a strict adherence to the Ward identity and the conservation of charge by construction, while our constraints encompass even-order radial moments of the renormalized vacuum polarization distribution and pointwise agreement with discrete samples of the vacuum polarization potential. The decomposition of the Coulomb, Breit and Uehling matrix elements in terms of one- and multi-centre contributions confirms that the Breit and Uehling parts are dominated by the one-centre contributions. In the case of the Breit interaction in closed-shell systems, the only non-vanishing contributions come from exchange interactions, and these are dominated by matrix elements involving inner-shell spinor amplitudes. This strongly suppresses all multi-centre contributions to the total energy from the Breit interaction. In the case of the Uehling interaction, the spatial extent of the Uehling potential (as well as the vacuum charge density that generates it) is limited to length scales that are of the order of the electron Compton wavelength $\lambdabar_{e}$. The integrand of the vacuum energy shift involves this effective potential and the one-particle density, which is dominated in the near-nuclear regions by the inner-shell electronic amplitudes. The precision with which one may evaluate the effects of vacuum polarization in molecules is mainly determined by the accuracy of the electron density in the neighbourhood of the nuclei, because we have established that the potential is readily represented to high accuracy as a Gaussian expansion. This renders the subsequent iterative inclusion of the Uehling potential in the self-consistent field procedure very efficient, because it uses elementary computational methods derived from the Gaussian product theorem.\\

In an extension of this study~\cite{fgq2024b}, we discuss an approach to the evaluation of higher-order QED effects in molecules, including the effects of the K\"{a}ll\'{e}n-Sabry and Wichmann-Kroll potentials, and the effects that arise due to coupling to virtual fields other than the electron-positron field. We find the general features described in this article are also pertinent to the study of higher-order QED effects.

\bibliographystyle{apsrev4-1}
\bibliography{biblio}

\begin{thebibliography}{20}%
\makeatletter
\providecommand \@ifxundefined [1]{%
 \@ifx{#1\undefined}
}%
\providecommand \@ifnum [1]{%
 \ifnum #1\expandafter \@firstoftwo
 \else \expandafter \@secondoftwo
 \fi
}%
\providecommand \@ifx [1]{%
 \ifx #1\expandafter \@firstoftwo
 \else \expandafter \@secondoftwo
 \fi
}%
\providecommand \natexlab [1]{#1}%
\providecommand \enquote  [1]{``#1''}%
\providecommand \bibnamefont  [1]{#1}%
\providecommand \bibfnamefont [1]{#1}%
\providecommand \citenamefont [1]{#1}%
\providecommand \href@noop [0]{\@secondoftwo}%
\providecommand \href [0]{\begingroup \@sanitize@url \@href}%
\providecommand \@href[1]{\@@startlink{#1}\@@href}%
\providecommand \@@href[1]{\endgroup#1\@@endlink}%
\providecommand \@sanitize@url [0]{\catcode `\\12\catcode `\$12\catcode
  `\&12\catcode `\#12\catcode `\^12\catcode `\_12\catcode `\%12\relax}%
\providecommand \@@startlink[1]{}%
\providecommand \@@endlink[0]{}%
\providecommand \url  [0]{\begingroup\@sanitize@url \@url }%
\providecommand \@url [1]{\endgroup\@href {#1}{\urlprefix }}%
\providecommand \urlprefix  [0]{URL }%
\providecommand \Eprint [0]{\href }%
\providecommand \doibase [0]{http://dx.doi.org/}%
\providecommand \selectlanguage [0]{\@gobble}%
\providecommand \bibinfo  [0]{\@secondoftwo}%
\providecommand \bibfield  [0]{\@secondoftwo}%
\providecommand \translation [1]{[#1]}%
\providecommand \BibitemOpen [0]{}%
\providecommand \bibitemStop [0]{}%
\providecommand \bibitemNoStop [0]{.\EOS\space}%
\providecommand \EOS [0]{\spacefactor3000\relax}%
\providecommand \BibitemShut  [1]{\csname bibitem#1\endcsname}%
\let\auto@bib@innerbib\@empty
\bibitem [{\citenamefont {Dyall}\ \emph {et~al.}(1989)\citenamefont {Dyall},
  \citenamefont {Grant}, \citenamefont {Johnson}, \citenamefont {Parpia},\ and\
  \citenamefont {Plummer}}]{dyall1989grasp}%
  \BibitemOpen
  \bibfield  {author} {\bibinfo {author} {\bibfnamefont {K.}~\bibnamefont
  {Dyall}}, \bibinfo {author} {\bibfnamefont {I.}~\bibnamefont {Grant}},
  \bibinfo {author} {\bibfnamefont {C.}~\bibnamefont {Johnson}}, \bibinfo
  {author} {\bibfnamefont {F.}~\bibnamefont {Parpia}}, \ and\ \bibinfo {author}
  {\bibfnamefont {E.}~\bibnamefont {Plummer}},\ }\href@noop {} {\bibfield
  {journal} {\bibinfo  {journal} {Computer Physics Communications}\ }\textbf
  {\bibinfo {volume} {55}},\ \bibinfo {pages} {425} (\bibinfo {year}
  {1989})}\BibitemShut {NoStop}%
\bibitem [{\citenamefont {Shabaev}\ \emph {et~al.}(2015)\citenamefont
  {Shabaev}, \citenamefont {Tupitsyn},\ and\ \citenamefont
  {Yerokhin}}]{shabaev2015qedmod}%
  \BibitemOpen
  \bibfield  {author} {\bibinfo {author} {\bibfnamefont {V.~M.}\ \bibnamefont
  {Shabaev}}, \bibinfo {author} {\bibfnamefont {I.~I.}\ \bibnamefont
  {Tupitsyn}}, \ and\ \bibinfo {author} {\bibfnamefont {V.~A.}\ \bibnamefont
  {Yerokhin}},\ }\href@noop {} {\bibfield  {journal} {\bibinfo  {journal}
  {Computer Physics Communications}\ }\textbf {\bibinfo {volume} {189}},\
  \bibinfo {pages} {175} (\bibinfo {year} {2015})}\BibitemShut {NoStop}%
\bibitem [{\citenamefont {Kahl}\ and\ \citenamefont
  {Berengut}(2019)}]{kahl2019ambit}%
  \BibitemOpen
  \bibfield  {author} {\bibinfo {author} {\bibfnamefont {E.~V.}\ \bibnamefont
  {Kahl}}\ and\ \bibinfo {author} {\bibfnamefont {J.~C.}\ \bibnamefont
  {Berengut}},\ }\href@noop {} {\bibfield  {journal} {\bibinfo  {journal}
  {Computer Physics Communications}\ }\textbf {\bibinfo {volume} {238}},\
  \bibinfo {pages} {232} (\bibinfo {year} {2019})}\BibitemShut {NoStop}%
\bibitem [{\citenamefont {Saue}\ \emph {et~al.}(2020)\citenamefont {Saue},
  \citenamefont {Bast}, \citenamefont {A.~S. P. Gomes H.~J}, \citenamefont
  {Visscher}, \citenamefont {Aucar}, \citenamefont {Remigio}, \citenamefont
  {Dyall}, \citenamefont {Eliav}, \citenamefont {Fasshauer} \emph
  {et~al.}}]{saue2020dirac}%
  \BibitemOpen
  \bibfield  {author} {\bibinfo {author} {\bibfnamefont {T.}~\bibnamefont
  {Saue}}, \bibinfo {author} {\bibfnamefont {R.}~\bibnamefont {Bast}}, \bibinfo
  {author} {\bibfnamefont {A.~J.}\ \bibnamefont {A.~S. P. Gomes H.~J}},
  \bibinfo {author} {\bibfnamefont {L.}~\bibnamefont {Visscher}}, \bibinfo
  {author} {\bibfnamefont {I.~A.}\ \bibnamefont {Aucar}}, \bibinfo {author}
  {\bibfnamefont {R.~D.}\ \bibnamefont {Remigio}}, \bibinfo {author}
  {\bibfnamefont {K.~G.}\ \bibnamefont {Dyall}}, \bibinfo {author}
  {\bibfnamefont {E.}~\bibnamefont {Eliav}}, \bibinfo {author} {\bibfnamefont
  {E.}~\bibnamefont {Fasshauer}},  \emph {et~al.},\ }\href@noop {} {\bibfield
  {journal} {\bibinfo  {journal} {The Journal of Chemical Physics}\ }\textbf
  {\bibinfo {volume} {152}},\ \bibinfo {pages} {204104} (\bibinfo {year}
  {2020})}\BibitemShut {NoStop}%
\bibitem [{\citenamefont {Flynn}\ \emph {et~al.}(2023)\citenamefont {Flynn},
  \citenamefont {Grant},\ and\ \citenamefont {Quiney}}]{FlynnGrantQuiney2023}%
  \BibitemOpen
  \bibfield  {author} {\bibinfo {author} {\bibfnamefont {D.~J.}\ \bibnamefont
  {Flynn}}, \bibinfo {author} {\bibfnamefont {I.~P.}\ \bibnamefont {Grant}}, \
  and\ \bibinfo {author} {\bibfnamefont {H.~M.}\ \bibnamefont {Quiney}},\
  }\href@noop {} {\bibfield  {journal} {\bibinfo  {journal} {Molecular
  Physics}\ ,\ \bibinfo {pages} {e2262610}} (\bibinfo {year}
  {2023})}\BibitemShut {NoStop}%
\bibitem [{\citenamefont {Grant}\ and\ \citenamefont
  {Quiney}(2022)}]{grant2022grasp}%
  \BibitemOpen
  \bibfield  {author} {\bibinfo {author} {\bibfnamefont {I.~P.}\ \bibnamefont
  {Grant}}\ and\ \bibinfo {author} {\bibfnamefont {H.~M.}\ \bibnamefont
  {Quiney}},\ }\href@noop {} {\bibfield  {journal} {\bibinfo  {journal}
  {Atoms}\ }\textbf {\bibinfo {volume} {10}},\ \bibinfo {pages} {108} (\bibinfo
  {year} {2022})}\BibitemShut {NoStop}%
\bibitem [{\citenamefont {Sunaga}\ \emph {et~al.}(2022)\citenamefont {Sunaga},
  \citenamefont {Salman},\ and\ \citenamefont {Saue}}]{sunaga20224}%
  \BibitemOpen
  \bibfield  {author} {\bibinfo {author} {\bibfnamefont {A.}~\bibnamefont
  {Sunaga}}, \bibinfo {author} {\bibfnamefont {M.}~\bibnamefont {Salman}}, \
  and\ \bibinfo {author} {\bibfnamefont {T.}~\bibnamefont {Saue}},\ }\href@noop
  {} {\bibfield  {journal} {\bibinfo  {journal} {The Journal of Chemical
  Physics}\ }\textbf {\bibinfo {volume} {157}},\ \bibinfo {pages} {164101}
  (\bibinfo {year} {2022})}\BibitemShut {NoStop}%
\bibitem [{\citenamefont {Shepler}\ \emph {et~al.}(2005)\citenamefont
  {Shepler}, \citenamefont {Balabanov},\ and\ \citenamefont
  {Peterson}}]{shepler2005ab}%
  \BibitemOpen
  \bibfield  {author} {\bibinfo {author} {\bibfnamefont {B.~C.}\ \bibnamefont
  {Shepler}}, \bibinfo {author} {\bibfnamefont {N.~B.}\ \bibnamefont
  {Balabanov}}, \ and\ \bibinfo {author} {\bibfnamefont {K.~A.}\ \bibnamefont
  {Peterson}},\ }\href@noop {} {\bibfield  {journal} {\bibinfo  {journal} {The
  Journal of Physical Chemistry A}\ }\textbf {\bibinfo {volume} {109}},\
  \bibinfo {pages} {10363} (\bibinfo {year} {2005})}\BibitemShut {NoStop}%
\bibitem [{\citenamefont {Shepler}\ \emph {et~al.}(2007)\citenamefont
  {Shepler}, \citenamefont {Balabanov},\ and\ \citenamefont
  {Peterson}}]{shepler2007hg}%
  \BibitemOpen
  \bibfield  {author} {\bibinfo {author} {\bibfnamefont {B.~C.}\ \bibnamefont
  {Shepler}}, \bibinfo {author} {\bibfnamefont {N.~B.}\ \bibnamefont
  {Balabanov}}, \ and\ \bibinfo {author} {\bibfnamefont {K.~A.}\ \bibnamefont
  {Peterson}},\ }\href@noop {} {\bibfield  {journal} {\bibinfo  {journal} {The
  Journal of chemical physics}\ }\textbf {\bibinfo {volume} {127}},\ \bibinfo
  {pages} {164304} (\bibinfo {year} {2007})}\BibitemShut {NoStop}%
\bibitem [{\citenamefont {Quiney}\ \emph {et~al.}(1997)\citenamefont {Quiney},
  \citenamefont {Skaane},\ and\ \citenamefont {Grant}}]{quiney1997}%
  \BibitemOpen
  \bibfield  {author} {\bibinfo {author} {\bibfnamefont {H.~M.}\ \bibnamefont
  {Quiney}}, \bibinfo {author} {\bibfnamefont {H.}~\bibnamefont {Skaane}}, \
  and\ \bibinfo {author} {\bibfnamefont {I.~P.}\ \bibnamefont {Grant}},\
  }\href@noop {} {\bibfield  {journal} {\bibinfo  {journal} {Journal of Physics
  B-Atomic Molecular and Optical Physics}\ }\textbf {\bibinfo {volume} {30}},\
  \bibinfo {pages} {L829} (\bibinfo {year} {1997})}\BibitemShut {NoStop}%
\bibitem [{\citenamefont {Grant}(2006)}]{grant2006}%
  \BibitemOpen
  \bibfield  {author} {\bibinfo {author} {\bibfnamefont {I.~P.}\ \bibnamefont
  {Grant}},\ }\href@noop {} {\emph {\bibinfo {title} {Relativistic Quantum
  Theory of Atoms and Molecules: Theory and Computation}}}\ (\bibinfo
  {publisher} {Springer-Verlag New York, Inc.},\ \bibinfo {year}
  {2006})\BibitemShut {NoStop}%
\bibitem [{\citenamefont {Andrae}(2000)}]{andrae2000}%
  \BibitemOpen
  \bibfield  {author} {\bibinfo {author} {\bibfnamefont {D.}~\bibnamefont
  {Andrae}},\ }\href@noop {} {\bibfield  {journal} {\bibinfo  {journal}
  {Physics Reports}\ }\textbf {\bibinfo {volume} {336}},\ \bibinfo {pages}
  {413} (\bibinfo {year} {2000})}\BibitemShut {NoStop}%
\bibitem [{\citenamefont {Abramowitz}\ and\ \citenamefont
  {Stegun}(1970)}]{abramowitz1970}%
  \BibitemOpen
  \bibfield  {author} {\bibinfo {author} {\bibfnamefont {M.}~\bibnamefont
  {Abramowitz}}\ and\ \bibinfo {author} {\bibfnamefont {I.}~\bibnamefont
  {Stegun}},\ }\href@noop {} {\emph {\bibinfo {title} {Handbook of Mathematical
  Functions}}}\ (\bibinfo  {publisher} {Dover Publishing Inc. New York},\
  \bibinfo {year} {1970})\BibitemShut {NoStop}%
\bibitem [{\citenamefont {Fullerton}\ and\ \citenamefont
  {Rinker}(1976)}]{fullerton1976}%
  \BibitemOpen
  \bibfield  {author} {\bibinfo {author} {\bibfnamefont {L.~W.}\ \bibnamefont
  {Fullerton}}\ and\ \bibinfo {author} {\bibfnamefont {G.~A.}\ \bibnamefont
  {Rinker}},\ }\href@noop {} {\bibfield  {journal} {\bibinfo  {journal}
  {Physical Review A}\ }\textbf {\bibinfo {volume} {13}},\ \bibinfo {pages}
  {1283} (\bibinfo {year} {1976})}\BibitemShut {NoStop}%
\bibitem [{\citenamefont {Glauber}\ \emph {et~al.}(1960)\citenamefont
  {Glauber}, \citenamefont {Rarita},\ and\ \citenamefont
  {Schwed}}]{glauber1960vacuum}%
  \BibitemOpen
  \bibfield  {author} {\bibinfo {author} {\bibfnamefont {R.}~\bibnamefont
  {Glauber}}, \bibinfo {author} {\bibfnamefont {W.}~\bibnamefont {Rarita}}, \
  and\ \bibinfo {author} {\bibfnamefont {P.}~\bibnamefont {Schwed}},\
  }\href@noop {} {\bibfield  {journal} {\bibinfo  {journal} {Physical Review}\
  }\textbf {\bibinfo {volume} {120}},\ \bibinfo {pages} {609} (\bibinfo {year}
  {1960})}\BibitemShut {NoStop}%
\bibitem [{\citenamefont {Huang}(1976)}]{huang1976calculation}%
  \BibitemOpen
  \bibfield  {author} {\bibinfo {author} {\bibfnamefont {K.-N.}\ \bibnamefont
  {Huang}},\ }\href@noop {} {\bibfield  {journal} {\bibinfo  {journal}
  {Physical Review A}\ }\textbf {\bibinfo {volume} {14}},\ \bibinfo {pages}
  {1311} (\bibinfo {year} {1976})}\BibitemShut {NoStop}%
\bibitem [{\citenamefont {Ward}(1950)}]{ward1950}%
  \BibitemOpen
  \bibfield  {author} {\bibinfo {author} {\bibfnamefont {J.~C.}\ \bibnamefont
  {Ward}},\ }\href@noop {} {\bibfield  {journal} {\bibinfo  {journal} {Physical
  Review}\ }\textbf {\bibinfo {volume} {78}},\ \bibinfo {pages} {182} (\bibinfo
  {year} {1950})}\BibitemShut {NoStop}%
\bibitem [{\citenamefont {Miranda}\ \emph {et~al.}(2012)\citenamefont
  {Miranda}, \citenamefont {Mendes}, \citenamefont {Gomes}, \citenamefont
  {Alves}, \citenamefont {Souza}, \citenamefont {Sambrano}, \citenamefont
  {Gargano},\ and\ \citenamefont {Macedo}}]{miranda2012}%
  \BibitemOpen
  \bibfield  {author} {\bibinfo {author} {\bibfnamefont {P.~S.}\ \bibnamefont
  {Miranda}}, \bibinfo {author} {\bibfnamefont {A.~P.}\ \bibnamefont {Mendes}},
  \bibinfo {author} {\bibfnamefont {J.~S.}\ \bibnamefont {Gomes}}, \bibinfo
  {author} {\bibfnamefont {C.~N.}\ \bibnamefont {Alves}}, \bibinfo {author}
  {\bibfnamefont {A.~R.}\ \bibnamefont {Souza}}, \bibinfo {author}
  {\bibfnamefont {J.~R.}\ \bibnamefont {Sambrano}}, \bibinfo {author}
  {\bibfnamefont {R.}~\bibnamefont {Gargano}}, \ and\ \bibinfo {author}
  {\bibfnamefont {L.~G.}\ \bibnamefont {Macedo}},\ }\href@noop {} {\bibfield
  {journal} {\bibinfo  {journal} {Journal of the Brazilian Chemical Society}\
  }\textbf {\bibinfo {volume} {23}},\ \bibinfo {pages} {1104} (\bibinfo {year}
  {2012})}\BibitemShut {NoStop}%
\bibitem [{\citenamefont {Dunning}(1989)}]{dunning1989}%
  \BibitemOpen
  \bibfield  {author} {\bibinfo {author} {\bibfnamefont {T.~H.}\ \bibnamefont
  {Dunning}},\ }\href@noop {} {\bibfield  {journal} {\bibinfo  {journal} {The
  Journal of Chemical Physics}\ }\textbf {\bibinfo {volume} {90}},\ \bibinfo
  {pages} {1007} (\bibinfo {year} {1989})}\BibitemShut {NoStop}%
\bibitem [{\citenamefont {Flynn}\ \emph {et~al.}(2024)\citenamefont {Flynn},
  \citenamefont {Grant},\ and\ \citenamefont {Quiney}}]{fgq2024b}%
  \BibitemOpen
  \bibfield  {author} {\bibinfo {author} {\bibfnamefont {D.~J.}\ \bibnamefont
  {Flynn}}, \bibinfo {author} {\bibfnamefont {I.~P.}\ \bibnamefont {Grant}}, \
  and\ \bibinfo {author} {\bibfnamefont {H.~M.}\ \bibnamefont {Quiney}},\
  }\href@noop {} {\  (\bibinfo {year} {in preparation, 2024})}\BibitemShut
  {NoStop}%
\end{thebibliography}%
\end{document}